\documentstyle[11pt]{article}

\begin{document}

\date{}
\title{A New Interpretation of Compensate Effect }
\author{{\ M.X Shao \thanks{{\protect\small E-mail: shaomingxue@hotmail.com}}}
, {Z. Zhao \thanks{{\protect\small E-mail: blackhole@ihw.com.cn}}} \\
{\small Department of Physics, Beijing Normal University, Beijing
100875, P.R.China.}} \maketitle
\begin{abstract}
A new interpretation of compensate effect is presented. The
Hawking effect in general space-time can be taken as a compensate
effect of the scale transformation  of coordinate time on the
horizon in generalized tortoise coordinates transformation. It is
proved that the Hawking temperature is the pure gauge of
compensate field in tortoise coordinates. This interpretation
does not refer to a zero-temperature space-time.

\end{abstract}

\vskip 0.6in

\section{Introduction}
Recently Zhao et al. developed an interpretation of Hawking
effect as a compensate effect of the coordinate scale
transformation\cite{hawking}\cite{D-R}\cite{sannan}\cite{zhao101}
\cite{zhao138}\cite{zhao139}\cite{zhao142}. In this understanding
the temperature  appears to be the $x^1$ component of a pure
gauge\cite{zhaobook}. It is of importance that the
zero-temperature space-time must be singled out. Otherwise an
uncertainty in this scheme  will arises. It is confident that all
is determined whenever the metric of space-time is given.
Therefore we look a scheme that the compensate effect is
determined by the space-time itself.  In section two a brief
review of Zhao's idea is given via a simple example. In section
three the temperature is presented to be $x^1$ component of a pure
gauge on the horizon in generalized tortoise coordinates. It is
proved for a general case and unnecessary to point out a
zero-temperature space-time.

\section{Zhao's Compensate Effect}
It is known that via the technique of conformal
flat\cite{zhao95}\cite{zhao96}\cite{zhao97}\cite{zhao98}\cite{zhao99}\cite{gr-qc0010079}
the line element in $x^0, x^1$ subspace is conformal to Minkowski
space near the event horizon. Zhao et al. analyze the conformal
factor and present the compensate effect. We now use Schwarzschild
metric in advanced Eddington-Finkelstein coordinate to go over
the skeleton.
\begin{equation}\label{com1}
ds^2=-(1-\frac{2m}{r})dv^2+2dvdr+r^2(d\theta^2+\sin^2\theta
d\varphi^2).
\end{equation}

In tortoise coordinates
\begin{equation}
r_*=r+\frac{1}{2\kappa}ln(\frac{r}{2m}-1),
\end{equation}
the line element can be rewritten as
\begin{equation}\label{com2}
ds^2=\Omega_2^2ds_2^2+r^2(d\theta^2+\sin^2\theta d\varphi^2),
\end{equation}
in which
\begin{equation}\label{com3}
\Omega_2^2=1-\frac{2m}{r}=\frac{e^{2\kappa(r_*-r)}}{2\kappa r}
\end{equation}
\begin{equation}\label{com4}
ds_2^2=-dv^2+2dvdr_*.
\end{equation}
 Introducing the generalized  null Kruskal coordinates
\begin{equation}\label{com5}
U=-\frac{1}{\kappa}e^{-\kappa u},~~V=\frac{1}{\kappa}e^{\kappa v},
\end{equation}
where the retarded Eddington-Finkelstein coordinate is defined as
$u=v-2r_*$. Its relation to the kruskal coordinates is
\begin{equation}\label{com6}
T=\frac{1}{2}(V+U),~~R=\frac{1}{2}(V-U).
\end{equation}
The line element Eq.(\ref{com2})can be written as
\begin{equation}\label{com7}
 ds^2=\Omega_1^2ds_1^2+r^2(d\theta^2+\sin^2\theta d\varphi^2),
\end{equation}
in which
\begin{equation}\label{com8}
\Omega_1^2=\frac{e^{-2\kappa r}}{2\kappa r}
\end{equation}
\begin{equation}\label{com9}
 ds_2^2=-dV^2+2dVdR.
\end{equation}

Since the line element keeps invariant, from
Eqs.(\ref{com2})(\ref{com7}), it is clear
\begin{equation}\label{com10}
ds_2^2=\frac{ds_1^2}{\Omega^2},~~\Omega^2=\frac{\Omega_2^2}{\Omega_1^2}.
\end{equation}
Eq.(\ref{com10})can be regarded as a conformal isometry.

Now consider a infinitesimal 2-dimensional vector $B_\mu$ whose
proper length is
\begin{equation}\label{com11}
  L=|ds|=\sqrt{B_\mu B^\mu}=\sqrt{g_{\mu\nu}B^\mu B^\nu},
  \mu,\nu=0,1
\end{equation}
whose coordinate lengths are defined respectively
\begin{equation}\label{com12}
  l_1=|ds_1|,~~l_2=|ds_2|.
\end{equation}
Obviously
\begin{equation}\label{com13}
  L=\Omega_1 l_1=\Omega_2 l_2.
\end{equation}
Since $L$ is invariant under parallel transfer in two dimensional
subspace ,
\begin{equation}\label{com14}
\delta ln l_1=-d ln \Omega_1,~~\delta ln l_2=-d ln\Omega_2
\end{equation}
is obtained. On the other hand, the contractions of the affine
connection of 2-dimensional subspace have the relations
\begin{equation}\label{com15}
 \Gamma^\alpha_{\alpha\mu}dx^\mu=d(ln\sqrt{-g_1})=2d(ln\Omega_1)
\end{equation}
\begin{equation}\label{com16}
\tilde{\Gamma}^\alpha_{\alpha\mu}dx^\mu=d(ln\sqrt{-g_2})=2d(ln\Omega_2)
\end{equation}
where $g_1$ and $g_2$ are the metric determinants. Define 1-form
$A_1$ and $A_2$
\begin{equation}\label{c17}
-A_1=\delta ln l_1,~~-A_2=\delta ln l_2,
\end{equation}
which rapidly results in
\begin{equation}\label{c18}
  A_1=\frac{1}{2}\Gamma^\alpha_{\alpha\mu},~~
  A_2=\frac{1}{2}\tilde{\Gamma}^\alpha_{\alpha\mu}.
\end{equation}
It can be find that $A_1$ and $A_2$ are respectively the relative
change rates of the coordinate lengths $l_1$ and $l_2$. The
coordinate length has the same scale as the coordinate time in
the two dimensional subspace, so both $A_1$ and $A_2$ reflect the
relative change rates of the coordinate times.  Since
Eq.(\ref{c18}) the $A$ transform as a connection in the scale
transformation
\begin{equation}\label{c19}
  A_2=A_1+d ln\Omega.
\end{equation}
The field strength 2-form
\begin{equation}\label{c20}
  F=0
\end{equation}
is immediately obtained since connection $A$ is an exact-form and
$dd=0$. It is easily obtained the property of pure gauge $A$
\begin{equation}\label{c21}
  \kappa=\frac{\partial ln\Omega}{\partial r_*}.
\end{equation}
It is the $x^1$ component of  connection that can be regarded as
Hawking temperature.

\section{Another Interpretation of Conformal Factor}

It is wonderful that the Hawking effect can be regarded as the
compensate effect of the coordinates. There is also disadvantage:
the zero-temperature space-time must be pointed out manually.
Otherwise this scheme can not determine the final result uniquely
since the method itself does not point out which space has the
temperature $T=\frac{\kappa}{2\pi K_B}$. Here we propose another
interpretation. The advantages of the new interpretation is that
the $\kappa$ is determined by the space-time itself and the
zero-temperature space-time is unnecessarily to be pointed out.
We directly give a general proof in advanced Eddington-Finkelstein
coordinates.

Consider the most general case that $\xi=\xi(x^0,x^2,x^3)$ in
Eddington coordinates. Tortoise coordinates transformation is
\begin{equation}
x_{*}^1=x^1+\frac1{2\kappa }ln(x^1-\xi ) \label{e4}
\end{equation}
with other components invariant.
\begin{equation}\label{c17}
dx^1_*=(1+\frac{1}{\epsilon})dx^1-
\frac{\xi^{\prime}_{\nu}}{\epsilon}dx^{\nu},
\end{equation}
in which $\epsilon=2\kappa(x^1-\xi)$ and
$\xi^{\prime}_\nu=\frac{\partial \xi}{\partial x^\nu}$.

 The metric is then obtained in terms of tortoise coordinates
\begin{equation}\label{c18}
ds^2=(\frac{\epsilon g_{11}\xi^{\prime}_0}{(1+\epsilon)^2}+
\frac{\epsilon g_{10}}{1+\epsilon})
[\frac{\frac{g_{11}}{(1+\epsilon)^2}\xi^{\prime}_0\xi^{\prime}_0
+\frac{2g_{10}}{1+\epsilon} \xi^{\prime}_0+g_{00}}{ \frac{\epsilon
g_{11}}{(1+\epsilon)^2}\xi^{\prime}_0+
g_{10}\frac{\epsilon}{1+\epsilon}}dx^0dx^0+2dx^0dx^1_*]+(others).
\end{equation}

The technique of conformally flat require around the horizon the
coefficient of $dx^0dx^0$ in $[~]$ in Eq.(\ref{c18}) being $-1$
\begin{equation}\label{c20}
\frac{\frac{g_{11}\xi^{\prime}_0\xi^{\prime}_0+2g_{10}\xi^{\prime}_0+g_{00}}
{\epsilon}+2g_{00}+2g_{10}\xi^{\prime}_0+\epsilon g_{00}}
{g_{10}+g_{11}\xi^{\prime}_0+\epsilon g_{10}}=-1.
\end{equation}

So  the line element in the two dimensional subspace $x^0, x^1_*$
is conformal to Minkowski space
\begin{equation}\label{com30}
  ds^2=\Omega^2(dx^0dx^0+2dx^0dx^1_*),
\end{equation}
where the conformal factor $\Omega$ is
\begin{equation}\label{i1}
\Omega^2=\frac{\epsilon g_{11}\xi^{\prime}_0}{(1+\epsilon)^2}+
\frac{\epsilon g_{10}}{1+\epsilon}.
\end{equation}

Putting Eqs.(\ref{c20})(\ref{i1}), the technique of conformally
 flatness at the event horizon obtains
 \begin{equation}\label{i2}
ds^2=\Omega^2|_{x^1\mapsto \xi}(-dx^0dx^0+2dx^0dx^1_*)+(other~~
terms),
\end{equation}
in which when $x^1$ approach the horizon $\xi$ only keeps the
first order of $\epsilon$, $\Omega^2$ equals to
\begin{equation}\label{i3}
\Omega^2|_{x^1\mapsto \xi}=\epsilon(g_{10}+g_{11}\xi^{\prime}_0).
\end{equation}
Finally the relation of $\kappa$ and when $x^1 \mapsto \xi$ the
$x^1$ component of connection $A$  is obtained as
\begin{equation}\label{i4}
  \frac{\partial ln\Omega}{\partial
  x^1_*}=\frac{1}{2}[\frac{\partial ln\epsilon}{\partial
  x^1_*}+\frac{\partial ln(g_{10}+g_{11}\xi^{\prime}_0)}{\partial
  x^1_*}]=\frac{1}{2}\frac{\partial ln\epsilon}{\partial
  x^1_*}=\kappa,
\end{equation}
where the condition is supposed that the components $g_{10} $ and
$g_{11}$ is not divergent on the event horizon $\xi$, i.e. their
derivatives with tortoise coordinate $x^1_*$ vanish when
$x^1\mapsto \xi$. In fact this hypothesis is not strong. For many
models of black holes in advanced Eddington-Finkelstein
coordinates, $g_{11}=0$ and $g_{10}$ is a constant.

It is emphasized that although  the Eq.(\ref{i4}) always holds,
generally speaking, for the most general horizon
$\xi=\xi(x^0,x^2,x^3)$ the $\kappa$ determined by this way is not
the one determined by the method of
Damour-Ruffini-Zhao\cite{zhao50}\cite{zhao78}\cite{zhao79}
\cite{zhao80}\cite{gr-qc0010078}


\begin{thebibliography}{99}
\bibitem{hawking} S.W.Hawking, Phys. Rev. Lett,26(1971)1344
\bibitem{D-R}  T. Damour and R.Ruffini. Phys Rev. D14(1976)332
\bibitem{sannan} S. Sannan. Gen. Rel. Grav, 20(1988)239
\bibitem{zhao101} Zhao Zheng and Liu Liao, Acta Physica Sinica
42(1993)1537
\bibitem{zhao138} Zhao Zheng and Li Zhong-heng, Acta Physica Sinica
45(1996)No.12.2091
\bibitem{zhao139} Zhao Zheng and Li Zhong-heng, Chin. Phys.
Bulletin, 40(1995)1951
\bibitem{zhao142}Ma Yong and Zhao Zheng, Chin. Phys.
Lett,13(1996)492
\bibitem{zhaobook}Zhao Zheng, Thermal Properties of Black Hole and
Singularity of spacetime, Beijing: the Beijing Normal University
Press,1999.
\bibitem{zhao95}Zhao Zheng and Huang Weihua, J.Beijing Normal Uni.(Science Version) 28(1992)317
\bibitem{zhao96}Ma Yong and Zhao Zheng, J.Beijing Normal Uni., Supplement(Science Version) 31(1999)70
\bibitem{zhao97}Zhao Zheng and Shen Chao, J.Beijing Normal Uni.(Science Version) 29(1993)194
\bibitem{zhao98}Zhao Zheng and Huang Weihua, J.Beijing Normal Uni.(Science Version) 29(1993)87
\bibitem{zhao99}Zhao Zheng and Huang Weihua, J.Beijing Normal Uni.(Science Version) 29(1993)90
\bibitem{gr-qc0010079}M.X. Shao and Zhao zheng, gr-qc/0010079(2000)
\bibitem{zhao50}Zhao Zheng and Dai Xianxin, Chin. Phys. Lett,
8(1991)548
\bibitem{zhao78}Zhao Zheng and Dai Xianxin, Chinese Science Bulletin 36(1991)1870
\bibitem{zhao79}Zhao Zheng and Dai Xianxin,Acta Physica Sinica
40(1995)23
\bibitem{zhao80}Dai Xianxin and Zhao Zheng, Acta Physica Sinica
41(1992)869
\bibitem{gr-qc0010078}M.X. Shao and Z.Zhao , gr-qc/0010078(2000)
\end{thebibliography}
\end{document}